# Māori algorithmic sovereignty: idea, principles, and use

*Paul T. Brown[1], Daniel Wilson[2], Kiri West[2], Kirita-Rose Escott[3], Kiya Basabas[4], Ben Ritchie[4], Danielle Lucas[4], Ivy Taia[1], Natalie Kusabs[1], Te Taka Keegan[1].*

[1] University of Waikato, Hamilton, NZ
[2] University of Auckland, Auckland, NZ
[3] Victoria University of Wellington, Wellington, NZ
[4] Nicholson Consulting, Wellington, NZ
Corresponding Author: Paul Brown (paul.brown@waikato.ac.nz)

**Abstract**

Due to the emergence of data-driven technologies in Aotearoa New Zealand that use Māori data, there is a need for values-based frameworks to guide thinking around balancing the tension between the opportunities these create, and the inherent risks that these technologies can impose. Algorithms can be framed as a particular use of data, therefore data frameworks that currently exist can be extended to include algorithms. Māori data sovereignty principles are well-known and are used by researchers and government agencies to guide the culturally appropriate use of Māori data. Extending these principles to fit the context of algorithms, and re-working the underlying sub-principles to address issues related to responsible algorithms from a Māori perspective leads to the Māori algorithmic sovereignty principles. We define this idea, present the updated principles and sub-principles, and highlight how these can be used to decolonise algorithms currently in use, and argue that these ideas could potentially be used to developed Indigenised algorithms.

**Keywords**: Indigenous; Māori; data sovereignty; data use; responsible algorithms.

## 1. Introduction

Artificial Intelligence (AI) technologies that involve data and algorithms are becoming more ubiquitous in decision-making processes (Olhede and Wolfe, 2018). These technologies are sold as solutions to biased human decision-making, and as a step towards wealthy societies, prosperity, and progress (Brynjolfsson and McAfee, 2014; Will *et al*., 2022). The positive intentions regarding the development and use of these technologies are usually genuine, however the outcomes may not be optimal from an equity perspective. A more critical look at the impact of these technologies shows that they have been used to generate wealth for large corporations (Lamdan, 2022), reduce labour costs and protect capital (Berardi, 2009; Meijas and Couldry, 2019). The impact of these uses has perpetuated and amplified historical injustices, particularly racial (Angwin *et al*., 2016; Beller, 2018; Benjamin, 2019; Checketts, 2022; Dressel and Farid, 2018), gendered (Buolamwini and Gebru, 2018; Lembrecht and Tucker, 2019), and economic injustice (Huws, 2014; Munn, 2017). Added to this is the significant energy costs associated with building and maintaining these systems which are detrimental to the environment (Henderson *et al*., 2020; Strubell *et al*., 2019). This creates additional burdens for marginalised communities that are least likely to realise the benefits, and most likely to be impacted by the harms of these technologies (Bender *et al*., 2021).

Indigenous voices are generally ignored in the process of building algorithms, from conception through to implementation and maintenance over its lifecycle. However, there is little to no hesitation in applying algorithms to these populations, sometimes for benevolent intentions, but often for exploitation and commercial profits (Munn, 2023; Walter and Anderson, 2013; Walter and Kukutai, 2018). Yet, Indigenous peoples have important and valuable perspectives to offer in this space. For example, the ideas of Indigenous data sovereignty (IDSov: Carroll *et al*., 2020; Rainie *et al*., 2019; Walter and Suina, 2019) have challenged the notion of data sovereignty by suggesting that data should be subject to the laws and governance structures of nations, including Indigenous nations, of those who the data is about, not just subject to the laws of the nation-state where the data lies (Kukutai and Taylor, 2016). This perspective offers the idea and a suite of methods for correcting power imbalances for Indigenous nations in a world where billions upon billions of bytes of data are collected, stored, bought, and sold amongst large corporations and governments for their own purposes.

The Māori data sovereignty (MDSov) principles is a good example of IDSov principles, specifically for the culturally appropriate governance and use of Māori data (Te Mana Raraunga, 2016), where Māori are the Indigenous people of Aotearoa New Zealand (hereafter referred to as Aotearoa, or NZ where appropriate). From these principles, several frameworks have been developed for specific uses, including Te Mana o te Raraunga for secondary use of Māori data in big data ecosystems (Hudson *et al.*, 2017), and Ngā Tikanga Paihere (Stats NZ, 2020a), a framework for culturally appropriate use of Māori data in the Integrated Data Infrastructure managed by Statistics NZ (Milne *et al.,* 2019). Recently developed works include the Māori Data Governance Model (Kukutai *et al.,* 2023a), and the Māori Data Sovereignty and Privacy Framework (Kukutai *et al.*, 2023b), both of which build upon the fundamental concepts and principles introduced in MDSov and apply them to the issues of data governance and data privacy respectively. In this paper, we introduce the idea of Māori algorithmic sovereignty (MASov), where algorithms can be understood as a particular use of data. We define a set of corresponding principles by extending the principles of MDSov to include appropriate use of algorithms that utilise Māori data, or that are applied to Māori individuals, communities, or environments that Māori have rights and/or interests in. The MASov principles are the starting point for the development of Māori tikanga (cultural)-based methods, frameworks, guidelines, or standards that can be used to assess existing algorithms that are applied to Māori, work towards decolonising existing algorithms, or developing indigenised algorithms so that they may produce fairer outcomes for Māori.

The structure of this paper is as follows; we define important terms in Section 2, including what we mean by algorithms, and define terms associated with algorithms. Section 3 introduces the idea of MASov and the corresponding principles and sub-principles that underpin MASov. We provide the reader with some historical context to justify the reasons why these principles are used and how they relate to the principles of responsible algorithms. Section 4 gives an example of using MASov principles to generate a framework to help assess an algorithm that produces biased outputs. We give our concluding remarks in Section 5 and discuss the transformational changes that may be possible for Māori as ideas such as MASov become more developed and used in practice.

## 2. Definitions and Terms

The term "algorithm" is broad and ambiguous and can mean different things under different contexts and perspectives. This is also true when defining some of the socio-technical terms often associated with algorithms, such as "bias", "fairness", and "transparency". This section provides the reader with some clarity about the terms we are using. We first distinguish between what we mean by computational algorithms and algorithmic systems, the latter of which we are interested in. Synonyms typically used for algorithmic systems include AI, AI systems, AI technologies, algorithms, automated decision-making processes, or models. We also provide some definitions regarding bias in relation to algorithms, definitions relating to Māori data, and definitions of terms typically used regarding responsible algorithm development, deployment, and use.

**Algorithms**

Dourish (2016) describes an algorithm from a computational perspective as "an abstract, formalised description of a computational procedure". The "computational algorithm" defined by Dourish, in practice, takes in a set of inputs as chosen by a user, and generates a set of outputs that is interpreted by a user (see Figure 1). This definition is too narrow for what we want to investigate. Not only do we want to investigate the inputs and outputs of the algorithm, but also the human decision-making that drives the process. In addition to this, we wish to elucidate the ways in which algorithms are shaped by society and vice versa. As such, we recognise a distinction here between computational algorithms as described by Dourish, and something more general, which we call an algorithmic system:

*Algorithmic System: An iterative decision-making process that is driven by humans, data, and computational algorithms.*

Therefore, when we talk about investigating algorithms, our intention is to investigate algorithmic systems as defined above. This is far broader than just looking solely at the computational algorithm. The scope of any analysis of algorithmic systems includes who is involved in all aspects of the development (from funders to designers and implementers), the motivations that drive the algorithm's existence, decisions regarding algorithm design, the inputs used, the outputs generated, the key decisions and policy that are made from the outputs, and the wider process including how the system is managed, monitored and maintained over time. For convenience, we will refer to algorithmic systems as "algorithms" and will distinguish between algorithms and computational algorithms to keep the terminology clear.

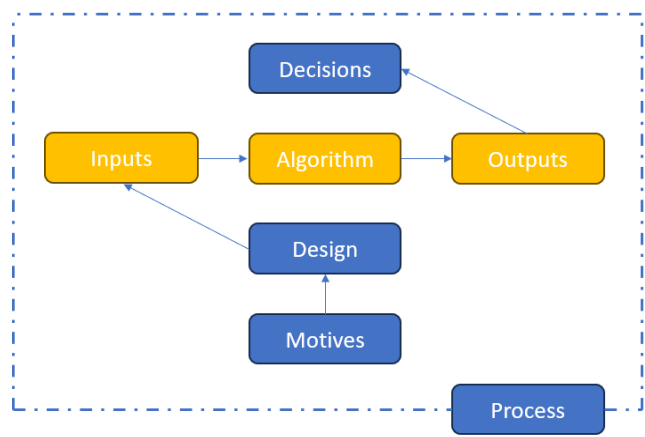

*Figure 1: A generic algorithmic system broken down into several foundational components.*

Figure 1 displays the structure of a generic algorithm, which is characterized by several different components. The yellow components are what we typically think of computational algorithms – a machine that takes inputs and generates outputs. However, there is a significant amount of human decision-making (blue components) that goes into this process, from the conceptualization of the algorithm through to implementation. Motives set the tone for the development, as this directly influences the aspirations the algorithm is constructed to achieve. The design component refers to how the algorithm is constructed to answer the motives, what data and variables are required for inputs, and how the data is to be collected. These inputs are fed into the algorithm component, which are the set of computational algorithms and specified model(s) that turn inputs into outputs. Once outputs are generated, there are interpretations and analyses of the outputs where new knowledge is gained before decisions about how the new knowledge will be used. The process component represents the decisions that are made throughout the lifecycle of the algorithm. An algorithm typically requires funding, and needs to be managed, maintained, monitored, and tested throughout its lifecycle. Note that the diagram in Figure 1 visually implies that the development of an algorithm is linear. However, we acknowledge in practice that this is most likely never the case! The dashed-dotted lines represent the iterative nature of the system, and the arrows represent the dependency structure of the system (e.g., the design depends on the motives that drive the algorithm development).

**Associated Terms**

In practice, algorithms are used to assist as a decision support, or decision-making tool, and provide economic value to the organisations that successfully implement them. However, there are ethical risks that can have detrimental impacts to the organisation (Someh, *et al.*, 2019) and society at large (Martin, 2019; O'Neil, 2016). One of the largest concerns is that the algorithms can replicate and perpetuate systemic biases (the inherent tendency of a process or system to favour certain outcomes) that exist or are inherent within society. This phenomenon is what is referred to as algorithmic bias, which occurs when the *outputs of an algorithm benefit or disadvantage certain individuals or groups over others without justification or reason for such unequal impacts* (Kordzadeh and Ghasemaghaei, 2022).

Systemic biases that algorithms perpetuate mainly impact marginalised populations, leading to the reinforcement of historical and current injustices, such as racial, gendered, and economic injustices. For the purposes of our analysis, the type of bias we are interested in is "colonising bias". We define colonising bias as follows:

*Colonising Bias: Prejudice or injustices against an indigenous group due to the effects of colonisation, that results in negative outcomes for that group*.

Colonising bias can be thought of as a particular type of racial bias. In the context of this article, we refer to colonising bias within algorithms, or colonising bias that occurs as a result of the outcomes of an algorithm *nested colonising bias*, and specifically focus on Māori as the indigenous people of Aotearoa, whose culture, traditions, and lives have been negatively affected by colonisation. We frame an algorithm as a use of data, thus if an algorithm uses Māori data, there are certain considerations that should be involved, and MDSov principles must be applied. Māori data is defined as *digital or digitisable information that is about, or generated from Māori people, Māori language, Māori culture, or resources and environments that Māori have rights and interests in* (Te Mana Raraunga, 2016). There are six principles generally associated with MDSov, Rangatiratanga (Authority), Whakapapa (Relationships), Whanaungatanga (Obligations), Kotahitanga (Collective Benefits), Manaakitanga (Reciprocity), and Kaitiakitanga (Guardianship). We expand more on what these ideas mean in the context of data and algorithms in the upcoming section, but interested readers can see Appendix 1 in Kukutai and Taylor (2016b) for further clarification.

As algorithms have become more prominent, private companies, research institutions, and government agencies around the world are racing to develop frameworks that contain principles that constitute the responsible development and use of algorithms. A scoping review surveyed 84 different documents containing ethical principles and guidelines to map the landscape of existing principles and to determine if a global convergence of certain principles was visible (Jobin *et al.*, 2019). The study found that the responsible algorithm principles of transparency, fairness, non-maleficence, responsibility, and privacy were cited by many of the documents as the principles most associated and important for responsible algorithms. Other principles highlighted were beneficence, freedom and autonomy, trust, sustainability, dignity, and solidarity. Whilst these principles do not necessarily have a strict definition in the context of algorithms, we have provided some general guidance on what these principles mean in Appendix 1, along with related terms.

3. **Māori Algorithmic Sovereignty**

MDSov is a particular case of IDSov and is the idea that Māori data should be subject to the laws and governance structures of Māori. Since algorithms depend on data, we take the perspective that MASov should be defined similarly. MASov is the idea that *algorithms that use Māori data, or that are applied to Māori individuals or collectives (including groups or organisations), or environments that Māori have rights and interests in, are subject to laws and governance structures of Māori.* As a broad and high-level idea, applying MASov gives Māori a way to meaningfully participate in all aspects of the development, deployment, and use of algorithms, protect Māori data and the information that stem from the outputs of algorithms, and partner with non-Māori to ensure use of algorithms uphold Māori rights, interests, and values.

Expanding on the idea of MASov, we specify a set of high-level principles and sub-principles that lay the foundations for what MASov is and what it might look like in practice. To understand how these principles are formulated, we provide some historical context before presenting the principles and sub-principles.

**Historical Context**

Māori are the original inhabitants of Aotearoa and settled the lands centuries prior to the arrival of the first Europeans (Walker *et al.*, 2017). In 1835, He Whakaputanga o te Rangatiratanga o Nu Tereni (the Declaration of Independence of the United Tribes of New Zealand) was signed as a formal deal between Māori and the British crown whereby mana (authority) and sovereign power of New Zealand rested with Māori, and foreign subjects could not make laws. The founding document of modern Aotearoa, Te Tiriti o Waitangi/The Treaty of Waitangi signed in 1840 between the Rangatira (tribal chiefs) of Aotearoa and representatives of the British Crown, established the foundations for the formation of a partnership between Māori and the Crown in Aotearoa. Generally speaking, the details of Te Tiriti are as follows; Article 1 establishes the authority of the Crown over its subjects through Kāwanatanga (loosely translated to mean governance). Article 2 gives recognition to the already established authority of Rangatira and grants continuation of their right of control over their taonga (treasured possessions, including objects, lands, and environment). Article 3 ensures individual citizenship rights and equality under the law. Both Articles 1 and 2 refer to the governing of two distinct populations and establish a framework for co-governance, whereas Article 3 speaks to individual citizen rights.

Since its signing, the importance and status of Te Tiriti has oscillated, but its role for Māori has always remained important (Hudson and Russel, 2009). What has not oscillated are past injustices, including treaty breaches, that Māori have been subject to historically and continue to this day (Belich, 1986; Moewaka Barnes and McCreanor, 2019). This has caused a large amount of distrust that Māori have toward Western colonial systems. Prominent Māori lawyer and scholar, Moana Jackson, notes that there is a tendency to historicise colonisation and its consequences as if it isn't a living, breathing reality (Jackson, 2019). The systems established in colonisation are still functioning today and continue to contribute to our contemporary struggles (Waziyatawin and Yellow Bird, 2005). Algorithms act as an extension of existing colonial infrastructure and as such, we see the extension of Indigenous distrust in these systems. Relatedly, there is a history of Indigenous communities, including Māori, being impacted negatively by quantitative research, which is generally imposed onto these communities, and have been used to reinforce negative stereotypes and reproduce deficit narratives (Smith, 2012; West et al., 2020; Walter and Anderson, 2013).

In the absence of a formal constitution, the Treaty of Waitangi Act 1975 was an important step in recognising the legal relevance of the Treaty of Waitangi and Te Tiriti as foundational documents of Aotearoa. The Act also led to the establishment of the Waitangi Tribunal, which allowed a legal pathway for Māori to redress historical grievances and injustices, such as land confiscation. In 1988, the NZ Royal Commission on Social Policy examined Te Tiriti, and the Treaty of Waitangi, and through their analysis identified the principles of (1) Partnership, (2) Protection, and (3) Participation. These principles imply that the crown has an obligation to recognise and empower Māori self-determination aspirations, and to protect Māori interests.

Currently, government agencies within Aotearoa are moving towards developing policies, legislation, and frameworks to fulfil their Tiriti obligations. For example, Manatū Hauora (NZ Ministry of Health) identified the principles of Tino Rangatiratanga (Māori self-determination), Equity, Active Protection, Options, and Partnership in a framework for delivering their services (Waitangi, 2019). The Data and Statistics Act 2022, whose purpose is to ensure high quality, impartial, and objective official statistics, acknowledges the Crown's responsibility to give effect to Te Tiriti and requires "the Statistician" to engage with Māori communities when collecting data and acknowledging Māori interests in data (see Sections 14(b) and 14(c)). This Act repealed the Statistics Act 1975, which made no mention or acknowledgement of Te Tiriti.

**Tiriti Principles and MDSov**

The principles outlining MDSov are fundamental concepts within Te Ao Māori (the Māori world), correspond to the principles of Te Tiriti (Partnership, Protection, and Participation), and are applicable to all Māori data. The establishment of the Kāwanatanga within the existing and established authority of Rangatira in Article 1 describes a partnership and a duty of care that each group has towards the other. The principles of Manaakitanga and Whakapapa speak to the principle of Partnership. Manaakitanga speaks to issues of respect for Māori data, and that free, prior, and informed consent underpins the collection, use and dissemination of Māori data. Whakapapa speaks to the acknowledgement of the genealogy of Māori data, the importance of data disaggregation, and decision-making around the use of Māori data to minimise future harms.

Article 2 of Te Tiriti acknowledges the existing authority of the Rangatira and their property rights over their taonga, speaking directly to the Tiriti principle of Protection. Both Dewes (2017) and Hudson *et al.*, (2016) argue that Māori data are observations of the world around them, and is a source of information about Māori people, Māori language, and Māori environments, which therefore constitutes a taonga. Recently, Waitangi Tribunal inquiries and reports such as WAI 262 (Waitangi, 2011) and WAI 2522 (Waitangi, 2016) formally recognised that all data have the potential to be taonga, reinforcing the assertion that the crown has a responsibility to protect Māori rights to data as affirmed by Te Tiriti. The Rangatiratanga principle speaks to Māori having control over their data and how it is used, the right to physically store Māori data in Aotearoa and the right to use Māori data in ways that empowers self-determination and furthers Māori aspirations. The Kaitiakitanga principle acknowledges that Māori have rights and obligations over their data, including the obligation to be responsible stewards over Māori data, that appropriate Māori ethical principles underpin the protection processes, and that Māori should decide what data is tapu (restricted) or noa (open).

The Te Tiriti principle of Participation relates to the equal individual rights for all citizens of Aotearoa and is enshrined in Article 3. The MDSov principles of Whanaungatanga and Kotahitanga stem from the Participation principle. Whanaungatanga speaks to the balancing of individual and collective rights, benefits and risks, and the accountabilities of individuals and organisations that are responsible for the generation, management, access etc. of Māori data, to the individuals, communities, and organisations of who the data derives from. Kotahitanga speaks to the idea that Māori should be able to derive individual and collective benefits from Māori data ecosystems, that capacity building and development of a Māori data workforce is needed, and that connections of other indigenous people with Māori must be encouraged.

**The MASov Principles**

The MDSov principles are Tiriti-centred and fit for the purpose of guiding appropriate governance and use of Māori data. Since algorithms are a specific use of data, we use the same six principles to underpin MASov. Many of the sub-principles remain, but the details of each are contextualised to the use of algorithms – including the use of Māori data, the computational algorithm, and the generated outputs of the system. Below are the MASov principles and sub-principles. Note that when referring to algorithms below, we mean all algorithms built that either (1) involve Māori data, (2) used to make decisions about Māori, (3) used to make decisions about environments that Māori have rights and interests in, or any combination of all three.

Rangatiratanga | Authority

1. Control – Māori have the right to control the development, and use of an algorithm, including (but not limited to) motives, design, choice of inputs, interpretation of outputs, maintenance, management, and deployment.
2. Jurisdiction - Decisions about the physical and virtual storage of the inputs and computational algorithms used, and the outputs generated from the algorithms shall enhance control for current and future generations. Whenever possible, the inputs and the outputs of the algorithms shall be stored in Aotearoa New Zealand.
3. Self Determination – Māori have the right to participate in the development and use of algorithms in a way that empowers sustainable self-determination and effective self-governance.

Whakapapa | Relationships

1. Transparency – Transparency in all aspects of the algorithm, including (but not limited to) who is involved, motivations, data and data provenance, outputs, management, maintenance, and deployment, should be clear prior to the application of the algorithm to ensure explainability.
2. Data Relationship – The use of Māori data throughout the algorithm process should be clear, and uphold the principles set out in MDSov.
3. Sustainability – It must be shown that the data and outputs used and generated from algorithms must provide long-term sustainable benefits to Māori, including environmental sustainability.

Whanaungatanga | Obligations

1. Balancing Rights – Individuals' rights, risks, and benefits in relation to the algorithms need to be balanced with the collectives they may be a part of.
2. Redress – Māori have the right to challenge the output or outcome of an algorithm if applied to them, and mechanisms for redress must be established in the process of algorithm development.
3. Accountability – Individuals and institutions that are responsible for the development of the algorithms are accountable to the Māori individuals and communities that the algorithm affects.

Kotahitanga | Collective Benefits

1. Benefit - Algorithms must be designed in ways that enable Māori to derive both individual and collective benefits, and to minimize harms.
2. Capacity Building – Individuals and institutions developing and using algorithms must include Māori in all parts of the process for meaningful partnership and to build capability for both Māori and non-Māori.
3. Solidarity – Māori must be supported to connect with other Indigenous groups for the purposes of sharing knowledge, ideas, and strategies regarding the development and use of algorithms. Where appropriate, Māori should also be supported to work with other groups that face discrimination from algorithms.

Manaakitanga | Reciprocity

1. Respect - The use of algorithms shall uphold the mana (respect) and dignity of Māori individuals and communities.
2. Privacy – Individual and collective privacy must be considered during the processes of data collection, storage, data re-use, and the dissemination of the outputs of the algorithm.
3. Consent – Any Māori community that an algorithm is applied to must give free, prior, and informed consent, for both the development and use of the system. This includes consents for data, outputs, and elements of the system that Māori control.

Kaitiakitanga | Guardianship

1. Protection – Inputs used in the algorithms and the resulting outputs must be treated in such a way that enables and reinforces the capacity of Māori to exercise kaitiakitanga over all components of the algorithm, including the inputs, outputs, and computational algorithms.
2. Ethics – Tikanga, kawa (protocols) and mātauranga (knowledge) shall underpin the protection, access, and use of the algorithms.
3. Restrictions – Māori shall decide how the inputs and outputs of the algorithms shall be considered tapu (restricted) or noa (accessible).

**More on the MASov Principles**

All MDSov sub-principles have been extended to include data, computational algorithms, and outputs, where necessary. There are issues with algorithms that are unique, and so we have changed and developed new sub-principles to address these issues. We note these changes below:

- Context (Whakapapa) has changed to Transparency, which acknowledges data genealogy and provenance, and extends this to all aspects of the algorithm to ensure that the algorithm is explainable.
- Data Disaggregation (Whakapapa) has changed to Data Relationship to ensure the way Māori data is used within an algorithm is tracked and that Māori data within an algorithm adheres to MDSov principles.
- Future Use (Whakapapa) has changed to Sustainability, to acknowledge that Māori have an interest in protecting their environments, and to ensure that the long-term benefits of an algorithm are sustainable.
- Redress (Whanaungatanga) was added to provide a mechanism for Māori to challenge the output or outcome of an algorithm that was applied to them. The right of redress is a Tiriti principle and is a widely discussed issue in responsible algorithms.
- Connect (Kotahitanga) has changed to Solidarity, to highlight that unity with all communities facing discrimination by algorithms (including Indigenous communities) is important and can benefit all.
- Privacy (Manaakitanga) has been added to ensure privacy of individuals and collectives is upheld and maintained through the algorithm process.

Table 1: The responsible algorithm principles alignment with MASov Principles.

| MASov Principle | Responsible Algorithm Principles |
|---|---|
| Rangatiratanga | Fairness and Justice, responsibility, beneficience, freedom, trust, dignity |
| Whakapapa | Transparency, responsibility, non-maleficience, beneficience, sustainability |
| Whanaungatanga | Transparency, fairness and justice, responsibility, trust, solidarity |
| Kotahitanga | Fairness and justice, non-maleficience, beneficience, dignity, solidarity |
| Manaakitanga | Responsibility, privacy, trust, dignity |
| Kaitiakitanga | Transparency, fairness and justice, responsibility, privacy, trust, sustainability, solidarity |

The set of principles underpinning responsible algorithm development, deployment, and use, as stated in Section 2 and Appendix 1, are noble and important principles to adhere to. The MASov principles and sub-principles touch on all the responsible algorithm principles. Moreover, they expand on issues specifically pertaining to Māori values such as the idea of stewardship of data and outputs (as opposed to ownership), highlight the importance of Māori culture, protocols, and knowledge when exercising stewardship over Māori data, the right of redress, and expand more on the issues of free, informed, and prior consent. Table 2 indicates which responsible algorithm principles correspond with the MASov principles.

## 4. Using MASov Principles

Government operationalised algorithms have been used in Aotearoa for some time and for many different applications (Stats NZ, 2018). The potential for algorithms and technologies causing harm through algorithmic bias has been recognised by the government. Several ethical frameworks for algorithm use have been developed along with the Algorithm Charter for Aotearoa NZ (Stats NZ, 2020b; Taylor Fry, 2021). Several issues have been raised with the Algorithm Charter from a Māori perspective. The charter mentions that it is a commitment by government agencies to reflect the principles of Te Tiriti. The charter notes that, under its commitment to partnership, it will embed a Māori perspective in the development and use of algorithms, yet nothing about the right for Māori to be active participants in this process, or provisions for the protection of Māori and Māori taonga. It also mentions the charters inability to "…fully consider important considerations, such as Māori data sovereignty, as these are complex and require separate consideration", though, as West *et al*., (2020) points out, MDSov should be integral to algorithm development.

There is a need to ensure that algorithms used or developed in Aotearoa are equitable and work for Māori. West *et al*., (2020) highlights the need for (a) the creation of a Māori values framework and tikanga guidelines to support automated decision-making design, development, use, and maintenance, (b) robust equity assessment protocols for algorithms, and (c) meaningful Māori participation in institutional algorithm self-assessment processes. The MASov principles defined in Section 3 are a solution to point (a) and can be used to develop frameworks to evaluate a particular algorithm and its potential for generating biased outputs, acting as a solution to point (b). The MASov principles may or may not increase meaningful Māori participation. However, it will provide non-Māori algorithm developers and users with some perspective of how algorithms should be applied to Māori communities.

In this section we put the MASov principles to use. We outline a strategy to investigate an algorithm that shows evidence of nested colonising bias through its outputs. We generate a framework for the assessment of an algorithm, where the MASov principles are used to generate in-depth questions that can investigate where, why, and how colonising biases may creep into the algorithm.

**Strategy**

In Section 2, we define a generic structure of an algorithmic system. We acknowledge that algorithms come in various forms, and our generic structure we defined may not encompass the important component that may be present in a particular algorithm. However, we will assume for the purposes of this exercise that this is the structure of an algorithm we wish to analyse. The values in Section 3 give us a guide as to the types of questions we may wish to ask at each component to analyse where, why, and how colonising bias may creep into the system. We could provide a fixed set of questions as a framework, but the idea of developing a more dynamic and living framework has several advantages. First, the problem with a generic structure and fixed set of questions is that it is not adaptive enough for the wide world of algorithmic constructions. Our previous work using a more fixed framework for analysis worked well for some algorithms but was significantly harder for others. Secondly, though it used MDSov principles in the structuring of the questions, it was not immediately apparent in some cases how these principles applied to algorithms. Having a dynamic framework coupled with well-defined Māori tikanga principles applied specifically for algorithms allow for a greater chance of detecting how and where nested colonising bias creeps into the system.

The full investigation of nested colonising bias requires a Kaupapa Māori mixed methods approach (Martel et al., 2022). Given we have a concerning algorithm that utilises Māori data, or is applied to Māori individuals or environments, the first requirement is to quantifiably assess whether the algorithm produces biased outcomes and determine the scale of the bias. If bias exists, we can qualitatively, and in a Māori way, assess the algorithm using the MASov principles to assess where, why, and how bias has crept into the system. If solutions can be found to tighten up the algorithm, another quantitative analysis to assess new outputs can be performed to quantify any existing biases and scale. The methods of the quantification of algorithmic bias are outside the scope of this paper, but for readers interested in case studies, see for example Bartlet *et al.,* (2021), Grother *et al.,* (2019), and Yesiler *et al.,* (2022).

The general strategy can be summarised in three steps: given we have an algorithmic system that produces biased outcomes for Māori:

1. Structure: define the key components of the algorithmic system and whether it contains nested colonising bias.
2. Framework: use the MASov principles to generate a framework to critically evaluate the components of the algorithm, providing insights to fix the system of its nested colonising biases.
3. Post-Analysis: perform an analysis where a comparison of bias in the fixed system and the old system is performed.

An example of generated table of questions in Appendix 2 is an example of a framework to assess each foundational component within an algorithm. The questions generated stem directly from the MASov principles, allowing Māori values to guide the deconstruction of the algorithm. In the tables, we have given a description of the component, prior information that may be needed, and a set of questions. The questions are structured under the headers of MASov principles. In practice, it may be more useful to ask questions pertaining to the sub-principles for a lower-level and detailed analysis.

### 5. Discussion – Towards Indigenised Algorithms

With algorithms becoming ever more ubiquitous, there is a growing need to ensure that these technologies are not used to perpetuate biases and cause harm to marginalised communities, including indigenous peoples. Algorithms currently deployed in Aotearoa have the potential of harming Māori, hence it is important that Māori can actively and meaningfully participate in the deployment, development and use of algorithms that are applied to them. In this paper we have presented the idea of MASov and defined its core principles. The principles are an extension of MDSov principles and are based on Te Ao Māori values and Te Tiriti principles. We also introduce the strategy of producing frameworks using these principles to investigate existing algorithms to understand, if evidence of colonising bias exists, how, where, and why it creeps into the system. Solutions may then be found to fix these issues, such that the algorithm can produce fairer results.

Algorithms already deployed are not embedded with Māori tikanga values and perspectives, and minimal consideration for Māori is usually given. Application of MASov principles for deconstructing and fixing existing algorithms are an incremental step towards change, but frustrating in the sense that the process of embedding tikanga values within algorithms should occur from the very start. What is required is transformational change in this space, where Indigenous values are embedded from the very start of development and continue throughout, from motives to decisions, and throughout the lifecycle of the algorithm. Figure 2 displays the idea of an indigenous algorithm, where an algorithm is housed in a structure founded on tikanga principles. The MASov principles are the tikanga values that could lay the foundation for the indigenous values, and whilst we recommend that they be used to retrofit existing algorithms, we would prefer that they be used as the foundational principles for the development of new, transformative, Indigenised algorithms.

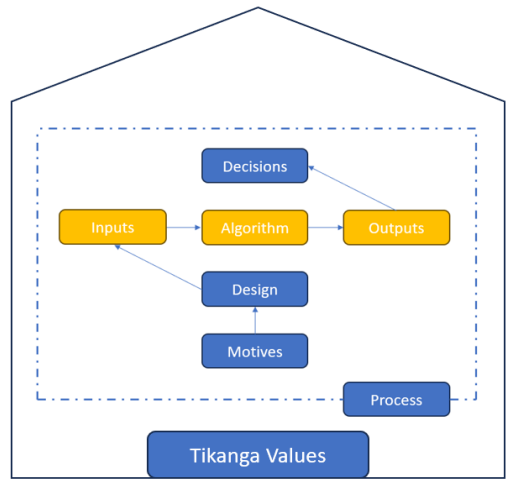

*Figure 2: An algorithm housed in a structure where tikanga values are at the foundation of the system.*

**Acknowledgements**

This work was funded by MBIE Tikanga in Technology (TinT), project no. 108303. Paul Brown's research was also funded by an MBIE Science Whitinga Fellowship, project no. 109120. The authors would like to thank Maui Hudson for his feedback, and the feedback from the participants of the 1st and 2nd Māori AI Wānanga and Workshops, both held at the University of Waikato on the 30-31st August 2022, and the 15th September 2023 respectively.

**Competing Interests**

The authors have no competing interests to declare.

# Appendix 1 – Principles for Responsible AI/Algorithms

Below are the principles of responsible algorithms as described in Jobin *et al.*, (2019).

| Principle | Meaning (in the context of algorithms) | Related principles |
|---|---|---|
| **Transparency** | Algorithms should be open, explainable, and explicit with regards to their purpose, development, use and maintenance. Efforts should be made to increase the explanability of how it works, interpretability of outputs, and understandability of the system. | Explainability, explicability, understandability, interpretability, communication, disclosure, non-opaque, showing. |
| **Fairness and Justice** | Outputs of algorithmic systems should be free of algorithmic bias, or at the very least, tested for bias and disclose findings. Purpose of the outputs must align with the ideas of fairness, equity and inclusion, and should be explainable enough so that the outputs can be challenged. | Consistency, inclusion, equality, equity, non-biased, non-discriminatory, diversity, plurality, accessibility, reversibility, remedy, redress, challenge, access, distribution. |
| **Non-Maleficence** | Algorithms should be safe and secure, and should not purposely cause foreseeable or unintentional harm (discrimination, physical harm, violation of privacy etc.). | Security, safety, harm minimisation, protection, precaution, prevention, integrity, non-subversion. |
| **Responsibility** | Use of algorithms should be done with integrity, and the allocation of responsibility, obligations, and legal liability should be clear in all parts of the process. A focus on the underlying sources of harm is necessary, as is the focus on diversity, inclusion, and participation of all relevant groups. | Responsible, accountability, liability, obligations, acting with integrity, participation. |
| **Privacy** | Algorithms require data, and privacy (typically presented in relation to data protection and security) is a value to uphold, and as a protected right! | Personal or Private information, anonymisation |
| **Beneficience** | Algorithms should be used in such a way that promotes human well-being and flourishing, peace, happiness, creation of socio-economic opportunites and prosperity. | Benefits, well-being, flourishing, peace, social good, common good |
| **Freedom** | Use of algorithms should promote freedoms, empowerment, autonomy, and self-determination through democratic means. There must be freedoms to withdraw consent, and individuals must be free from manipulation, surveillance, or technological experimentation. | Autonomy, consent, choice, self-determination, liberty, empowerment |
| **Trust** | Trust in Algorithms refers to the algorithm having a noble purpose, used by trustworthy individuals and organisations, built on good design principles, and should aspire to gain the trust of stakeholders. | Trust, Purpose |
| **Sustainability** | Development and deployment of algorithms should consider the protection of the environment, improving Earth's ecosystem and biodiversity, contribution to fair and equitable societies, and the promotion of peace. | Environment (nature), energy, resources (energy) |
| **Dignity** | Algorithms should uphold human rights. It should not diminish or destroy, but preserve or increase human dignity. | Human dignity |
| **Solidarity** | Algorithms have a large implication on the labour market, so benefits of the algorithm should uphold strong safety nets, wealth generated should be redistributed to those whose labour has been taken through automation. | Solidarity, social security (welfare), cohesion, redistribution of benefits |

## Appendix 2 – Generated Framework Using MASov Principles

Below are generated questions using the MASov principles, for the algorithm components of process, motives, inputs, and outputs.

| | |
|---|---|
| **Process** | The Process component looks at the algorithm in its entirety. Important information to extract here are the people/organisations involved in the creation, development, maintenance, ownership, and funding. It is also important to understand the intent of the algorithm, how the data is being protected, and how the system will be maintained throughout its use.<br><br>Questions to investigate the algorithm come straight from the MASov principles, which will provide a high-level look at the system. |
| Example Questions | Rangatiratanga: What controls do Māori have in all stages of the development of the algorithm?<br><br>Whakapapa: Has this algorithmic system been used previously on Māori or other indigenous communities? For what purpose?<br><br>Whanaungatanga: How are individuals and institutions involved in the development and use of the system accountable to Māori?<br><br>Kotahitanga: What strategies for capacity building are there to ensure technical literacy in the Māori communities to which the algorithm applies?<br><br>Manaakitanga: Have the necessary Māori individuals and communities given free, informed, and prior consent for their data to be used in the algorithm?<br><br>Kaitiakitanga: Do Māori ethics underpin the protection, access, and use of the algorithm? |
| **Motives** | The Motives component looks at understanding what problems the algorithm is solving (or goals it is trying to achieve) and asks questions involving what the problem/goals are, and who is involved when defining the problems/goals, and where/if Māori consultation has been sought. This helps understand if an algorithmic system is the correct tool for solving the problem/achieving the goal.<br><br>Preliminary steps before analysis would be to understand the underlying motivations for the algorithm, and whose motivations are driving the process, what the system is trying to achieve, who is involved in the process, and if Māori in any way have been involved in defining the motivations and in what capacity? |
| Example Questions | Rangatiratanga: Do the motivations/purpose of the algorithm further Māori collective aspirations?<br><br>Whakapapa: Are the motivations underlying the use of the algorithm clear in providing future benefit to Māori<br><br>Whanaungatanga: Which individuals and institutions have defined the motivations of the algorithm, and what are their obligations to Māori?<br><br>Kotahitanga: What harms and benefits do the motivations provide for Māori?<br><br>Manaakitanga: Do the motivations uphold and maintain dignity for Māori individuals and communities?<br><br>Kaitiakitanga: Do Māori have the right to change the motivations if tikanga values are not involved in the construction of the motivations? |
| **Inputs** | The Inputs component looks specifically at what data, variables, and other inputs (such as weights, priors, model specifications) have been chosen to be used in the computational algorithm. The inputs all depend on the decisions made and the data collected throughout the process so far, so it is important to ask questions regarding the consistency of the process thus far as we approach this pivotal step of the process.<br><br>Prior to analysis, it is important to understand the technical components of the process, including understanding what variables have been defined for the eventual model, what data has been used and whether it is sufficient, details surrounding the model, and the algorithms that could be considered useful to run with the data available. |
| Example Questions | Rangatiratanga: What controls do Māori have to determine what inputs are tapū (closed) or noa (open)?<br><br>Whakapapa: Do the inputs used align with the (Māori) motivations of the algorithm?<br><br>Whanaungatanga: Who (individuals/institutions) is responsible for the protection of inputs (Māori data)?<br><br>Kotahitanga: What are the strategies to build technical capacity and knowledge for Māori issues surrounding the protection of inputs?<br><br>Manaakitanga: Have appropriate Māori communities given consent for the application of their data being used in the algorithm?<br><br>Kaitiakitanga: Do the inputs have the necessary protocols in place for protection and security? |
| **Outputs** | Once inputs are chosen, they are plugged into the computational algorithms, and outputs are generated. The Outputs component looks at how the outputs are interpreted and communicated to decision-makers. Other important aspects involve access to outputs, benefit-sharing, capacity building, and if/how Māori are involved with the interpretation of outputs.<br><br>Important things to understand is who is analysing and interpreting the outputs, the quality of the outputs, and who decides if outputs are correct or relevant to the motivations of the algorithm. Since outputs are newly generated knowledge created from inputs, it is important that all outputs that are about Māori are treated with the same care and respect as Māori data. |
| Example Questions | Rangatiratanga: Do the outputs and the analysis and interpretation of the outputs contribute to Māori self-determination and aspirations?<br><br>Whakapapa: Are outputs about Māori consistent with the inputs (specifically Māori data)?<br><br>Whanaungatanga: What are the obligations of the individuals and institutions that generate new outputs, to the Māori individuals and communities that the outputs describe?<br><br>Kotahitanga: Have the outputs been interpreted from the correct Māori lens?<br><br>Manaakitanga: Do the findings of the outputs, and the analysis and interpretation of the outputs uphold Māori dignity?<br><br>Kaitiakitanga: Are the outputs about Māori treated the same as Māori data with respect to controls and protections? |